\documentclass[sigconf,natbib=true]{acmart}

\usepackage{graphicx}
\usepackage{soul}
\usepackage[utf8]{inputenc} 
\usepackage[T1]{fontenc}    
\usepackage{hyperref}
\usepackage{url}            
\usepackage{booktabs}       
\usepackage{amsfonts}       
\usepackage{nicefrac}       
\usepackage{microtype}      
\usepackage{xcolor}         

\usepackage{amsmath}
\usepackage{nccmath}
\usepackage{setspace}
\usepackage{wrapfig}
\usepackage{caption}
\usepackage{subcaption}
\usepackage{amssymb}
\usepackage{bbding}
\usepackage{multirow}
\usepackage{algorithm}
\usepackage{algpseudocode}
\usepackage{listings}
\usepackage{natbib}
\usepackage{minitoc}
\usepackage{fancybox}
\usepackage{enumitem}
\usepackage{float}
\usepackage{lipsum}
\usepackage{tablefootnote}

\usepackage[nameinlink]{cleveref}
\usepackage[titletoc]{appendix}

\theoremstyle{plain}

\newtheorem{myTheo}{Theorem}

\AtBeginDocument{%
  \providecommand\BibTeX{{%
    \normalfont B\kern-0.5em{\scshape i\kern-0.25em b}\kern-0.8em\TeX}}}

\begin{document}

\title{Confidence Ranking for CTR Prediction}

\author{Jian Zhu, Congcong Liu, Pei Wang, Xiwei Zhao, Zhangang Lin, Jingping Shao }

\affiliation{%
  \institution{JD.com}
  \country{Beijing, China}
  }
  
\email{ {zhujian146, liucongcong25, wangpei959,zhaoxiwei,linzhangang,shaojingping}@jd.com }
\renewcommand{\shortauthors}{Zhu and Liu, et al.}

\begin{abstract}
Model evolution and constant availability of data are two common phenomenon in large-scale real-world machine learning application, e.g. ads and recommendation system.
To adapt, real-world system typically \textit{retrain} with all available data and \textit{online learn} with recent available data to update the models periodically with the goal of better serving performance.
However, if model and data evolution results in a vastly different training manner, it may induce negative impact on online A/B platform.
In this paper, we propose a novel framework, named \textit{Confidence Ranking}, which designs the optimization objective as a ranking function with two different models.
Our confidence ranking loss allows direct optimization of the logits output for different convex surrogate function of metrics, e.g. \textit{AUC} and \textit{Accuracy} depending on the target task and dataset.
Armed with our proposed methods, our experiments show that the confidence ranking loss can outperform all baselines on CTR prediction of public and industrial datasets. 
This framework has been deployed in the ad system of JD.com to serve the main traffic in the fine-rank stage.

\end{abstract}

\ccsdesc[500]{Information systems~Retrieval models and ranking}
 \ccsdesc[500]{Information systems~Online advertising}
 \ccsdesc[500]{Information systems~Recommender systems}

\keywords{Click-Through Rate Prediction; Loss function; Deep learning}

 \maketitle

\section{Introduction}
\label{introduction}
To alleviate the discrepancy between delayed training and test distribution, typically,
the real-world ads and recommendation system widely operate machine learning pipeline as follows: (1) collect user-clicks periodical data (every $\Delta =24$ hours for daily data and $\Delta < 10$ minutes for online data); (2) train the model on collected data then deploy the model for serving until next new retrained model is produced. Due to the non-stationary data distribution and constant model evolution, the periodic retraining methodology motivate fast adaption and better generalization for approximating the recent decision boundary. While retraining with new data and all data does make the online model more current and confident, the improvement only originates from the alleviation of train-test discrepancy or promotion of model capability. We investigate if modeling previous prediction(confidence) distribution can improve the performance even further. For short, as we train the model at a time $t$ with data $\mathcal{D}^t$, we only know that the model will be optimized in this period while ignoring how the data are produced and influenced by the online deployed model. \textbf{Our goal in this paper is to improve test-set performance of retrained model compared to the online deployed model for CTR prediction.}

\par To improve model performance, a common solution knowledge distillation \citep{hinton2015distilling}, which involves having a \textit{teacher} model and mixing its predictions with original labels has proved to be a useful tool in deep learning. Recently, distillation-based works\cite{tang2018ranking,kang2021topology,hofstatter2020improving} have been widely used in recommendation system for training efficient and effective student models. Despite its success, this methodology requires the teacher model performs better than students while the divergence of model capacity is not large \cite{dao2021knowledge}. In this way, knowledge distillation may constrain its power when applied in CTR prediction pipeline because the capacity of retrained model maybe not weaker than the online model. Thus, an important question remains open regarding machine learning pipeline of real-world system:

\shadowbox{
	\parbox{73mm}{
		\centering
		\textbf{\textit{How can we train a model better than the online deployed model?}
	}}
}

\par In this paper, we provide an affirmative answer to this question. Our solution is to optimize a novel framework of loss function for machine learning application (MLA) instead of cross-entropy optimization. In particular, we choose to maximize the ranking score between base model and retrained model for CTR prediction. There are several benefits of maximizing the ranking score over minimizing the cross-entropy loss. First, maximizing the ranking score is naturally suitable for classification task in real-world CTR prediction where train-test distribution is nearly identically independent distributed. Directly maximizing the ranking score of different models can approximate improving the model performance relatively. Second, this framework is more suitable for handling various model complexity since maximizing ranking score aims to learn better decision boundary compared to baseline. The foremost challenge in this paper is to determine the surrogate loss for this setting. In our study, a naive approach of exploiting ranking-based pair-wise surrogate loss can be efficiently optimized for various metrics(e.g. AUC and Accuracy). Our contribution can be summarized as: (1) Our proposed loss framework achieves state-of-the-art performance for CTR prediction and we have deployed it to real-world ads system on CTR in the fine-rank stage until now. (2) We give theoretical and experimental analysis on our proposed confidence ranking loss showing the superiority over distillation.

\section{Methods}

\subsection{Preliminaries}

\begin{figure}[t]
	\centering
	\includegraphics[width=0.45\textwidth]{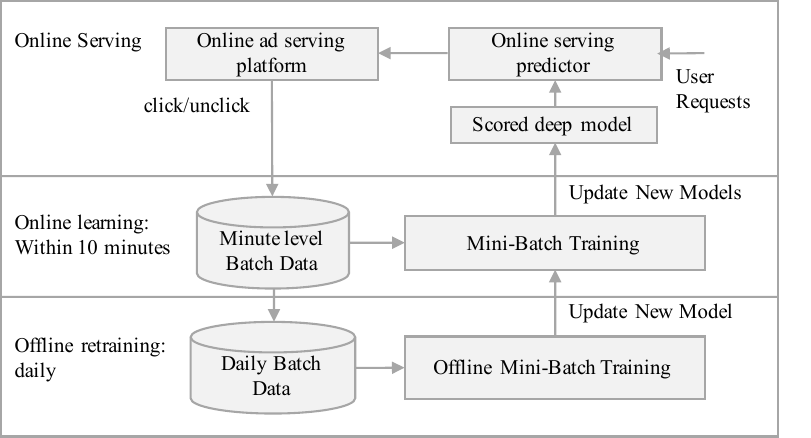}
	\caption{brief system overview of CTR prediction pipeline.}
	\label{fig:system}	
\end{figure}

Formally, as illustrated in Figure \ref{fig:system}, a CTR prediction pipeline in real-world system can be split to three parts: \textbf{(1)} Offline training: given old dataset $\mathcal{D}_{old}$ that consists of continuous \textit{T} days data with corresponding input sample $x$ ad ground truth labels $y$, we build a machine learning model $f$  with parameters $\theta$ for the aim to optimize in a sequential manner. We note the output logits with $z \triangleq h(x;\theta)$, ground truth $y$ and the corresponding  probability with $f(x;\theta) \triangleq {\rm sigmoid}(h(x;\theta))$. The goal of first part is to optimize $f$ on the $\mathcal{D}_{old}$ with cross-entropy optimization:

\begin{equation}
	\mathop{\arg\min}\limits_{\theta} \mathcal{L}_{old}, \quad  {\rm where}   \quad  \mathcal{L}_{old} \triangleq \mathbb{E}_{(x,y) \sim \mathcal{D}_{old}}[\ell(y, f(x;\theta))]
\end{equation}
\textbf{(2)} Online serving: after the loss $\mathcal{L}_{old}$  converge smaller than $\delta$, where $\delta$ is specified by cross-validation on $\mathcal{D}_{old}$, we deploy the machine learning model $f_{\theta}$ on the real-world system for the aim to serve the new arriving dataset $\mathcal{D}_{new}$ where the ground truth $y_{new}$ is unknown until the user clicks the item or leave the browser. Until now, we have demonstrate the pipeline of real world deployment applications.
\textbf{(3)} Online learning: As we get new arriving data, we retrain previous deployed model $f_{old}$ in order to overcome the challenge of inconsistency between $\mathcal{D}_{old}$ and $\mathcal{D}_{new}$ before online serving with new model $f_{new}$. Benefited from mitigation on distributional drift, the strategy of online learning with recent data can efficiently improve the generalization on the next serving stage under the assumption that recent activities reflect the users' evolving interest. Despite effectiveness of the pipeline, this strategy does not take previous model's outputs as auxiliary information which covers the underlying relationship between prediction and ground truth. It keeps unknown that how a vastly different training manner induced by model and data evolution impact online performance.
\par As we open an important question of how to learn better in both stages in our system, we would ideally like to define the concept "better" as difference of the metric score generated by $f$ and $f_{online}$ where $f$ and $f_{online}$ is the models that we currently train and deploy on the online serving platform respectively. Given a metric function $\mathcal{M}(y,\hat{y})$, where $y$ and $\hat{y}$ is the ground truth labels and predicted values, we then seek to minimize the classification risk for $f$, subject to the metric score generated by $f$ should be better than $f_{online}$:
\begin{equation}
    \label{eq:metric_ranking}
    \mathop{\arg\min}\limits_{\theta} \mathcal{L} \quad s.t. \quad \mathcal{C}(f)>0 
\end{equation}
where $C(f) \triangleq \mathcal{M}(y, f(x))-\mathcal{M}(y, f_{online}(x)) $.
In this way, we can design algorithm in the mini-batch training strategy in both stages as shown in Figure \ref{fig:system} with extra online model predictions. However, directly optimizing the metric score is not differentiate for deep models. For the purpose of designing a tractable algorithm, we will instead work with softer notion of "better", which evaluates the divergence of their metric scores.

\subsection{Confidence Ranking}
Metric loss typically are calculated by 0-1 loss. For example, accuracy and AUC of a mini-batch samples can be defined by $acc=\frac{1}{n}\sum_{i=1}^{n}\mathbb{I}[y_i==\hat{y}_i]$ and $auc=\frac{1}{nm}\sum_{i=1}^{n}\sum_{j=1}^{m}\mathbb{I}[\hat{y}_i^+ > \hat{y}_j^-]$ respectively, where $\mathbb{I}$ is indicator function. To ease optimization, researchers resort to surrogate score function when maximizing accuracy and auc \cite{mohri2018foundations}. Thus, we can devise confidence-ranking loss to directly employ previous learned knowledge. 

\par \textbf{Confidence Ranking (CR) for Accuracy.} As we want to achieve better accuracy, the expected metric objective can be defined as $C_{acc}(f)=\frac{1}{n}\sum_{i=1}^{n}\mathbb{I}[y_i f(x_i)>y_i f_{online}(x_i)]$. 
As we want to maximize this objective, we only need to induce a surrogate loss function to rank the point-wise model outputs which can be defined as:

\begin{equation} \label{risk_form}
	\ell_{CR}(f) \triangleq \mathbb{E}_{\{x,y\} \sim \mathcal{D}} \left[(\phi_{y}(f(x) - f_{online}(x)) \right] 
\end{equation}
where we only consider scoring functions $\phi_y$ that are \textit{strictly proper} \cite{gneiting2007strictly} (e.g. logistic rank loss $\phi_y(u,v)=log(1+exp^{-(u-v)})$ and square loss $\phi_y(u,v)=(1-(u-v))^2$). For simplicity, in this work we only consider the logistic loss function which can be defined as: 

\begin{equation}\label{point_log_cr}
	\ell_{CR} = \frac{1}{n}\sum_{i=1}^n y_i log(1+exp^{-(u-v)}) + (1-y_i)log(1+exp^{(u-v)})
\end{equation}
where $u$ and $v$ are $h(x_i)$ and $h_{online}(x_i)$ respectively.

\par \textbf{Relational Confidence Ranking (RCR) for AUC:} The point-wise loss that rank the output of current model with online deployed one ensures the network gradually perform better. To further improve the bipartite ranking performance of binary classification, we follow \cite{freund2003efficient} in optimizing bipartite ranking performance. As we want to achieve better bipartite ranking performance, the expected metric objective can be defined as $C_{auc}(f)=\frac{1}{nm}\sum_{i=1}^{n}\sum_{j=1}^{m}\mathbb{I}[(f(x_i^+)-f(x_j^-))>(f_{online}(x_i^+)-f_{online}(x_j^-))]$ where $x_i^+$ and $x_j^-$ are $i$-th positive sample and $j$-th negative sample. Thus, we define \textit{relational confidence ranking} risk as:

\begin{equation}
	\ell_{RCR}(f) \triangleq \mathbb{E}_{ \{x^+, x^-\}  \sim \{\mathcal{P}^+, \mathcal{P}^- \} } \left[ \phi(d_f(x^+, x^-) - d_{f_{online}}(x^+, x^-)) \right]  
\end{equation}
where function $d_f(x, z)=f(x)-f(z)$  performs calculating distance of different samples $x$ and $z$, and $\mathcal{P}^+$ and $\mathcal{P}^- $ are the positive and negative classes respectively. Similar to point-wise confidence-ranking loss, we select logistic loss function as our scoring function:
\begin{equation}
	\ell_{RCR} =  \frac{1}{nm}\sum_{i=1}^n \sum_{j=1}^m log(1+exp^{-(u-v)})  
\end{equation}
where $u$ and $v$ are $d_h(x_i^+,x_j^-)$ and $d_{h_{online}}(x_i^+,x_j^-)$ respectively.

\par \textbf{Training with Confidence Ranking.} During training, multiple confidence ranking loss function, including the proposed point-wise CR loss and relational CR loss can be either alone or together with task-specific loss functions, e.g. cross-entropy for classification. Therefore, the final objective is defined as:
\begin{equation}
    \ell_{ce}+\lambda_{CR_{acc}} \ell_{CR} + \lambda_{RCR_{auc}} \ell_{RCR}
\end{equation}
where $\ell_{ce}$ is a cross-entropy loss in CTR prediction, $\ell_{CR}$ and $\ell_{RCR}$ are the point-wise and relational confidence ranking loss respectively, and $\lambda_{CR_{acc}}$ and $\lambda_{RCR_{auc}}$ are tunable hyperparameters to control the loss terms. For sampling tuples of pos/neg samples in the proposed relational confidence ranking loss, we simply use all possible pairs in a given mini-batch.

\begin{myTheo} 
    \label{thm:bias_variance_cr}
    \begin{sloppypar}
    (\textbf{Bias-Variance bound for confidence ranking})\label{prop_cr}
    Pick any convex loss $\ell$. Suppose we have a teacher model $p^t$ with corresponding empirical confidence ranking  risk $\widehat{R}(f)=\dfrac{1}{N}\sum_{n \in N} y(x_n) \ell(\text{f}(x_n)-f_t(x_n)))$ and population risk $R(f)=\mathbb{E}_x \left[ p^*(x) \ell(f(x))\right] $  where $f_t(x_n)$ is the teacher output. For any predictor $f$: $\mathcal X \rightarrow \mathbb R^L$,
    	\begin{equation}
    		\label{bound_cr}
    		\mathbb E \left[ (\widehat{R}(f) - R(f))^2 \right]  \leq  \mathbb{E} \left[ (R(f_t))^2 \right] 
    	\end{equation}
    \end{sloppypar}
\end{myTheo}

We have stated a statistical perspective on confidence ranking, resting on the observation that confidence ranking offers a bound which always approximating Bayes probabilities based on the performance of teacher model. However, this bound is not well qualified on deep learning architecture and may be loose and unstable for real-world application especially for the logistic confidence ranking loss. We note the comprehensive bound of confidence ranking requires specifying necessary conditions. Nonetheless, this qualitative bound can still hold majority conditions in practice.

\begin{table}
		\centering
		\small
		\caption{The statistic of CTR prediction datasets}
		\begin{tabular}{lccccc}
			Datasets & Users & Items & Fields & Feature size & Instances  \\
			\toprule
			Avazu & N/A & N/A & 22 & 2018012 & 40428967 \\
			Avito 		 & 	3163597  &  28529 & 16 & 3419165 		& 190107687  \\     
			Industrial & N/A 			& N/A 		& 59 & N/A 					& 12 Billion \\
			\bottomrule
		\end{tabular}
		
		\label{tab:ctr_dataset}
\end{table}

\begin{table}
	\centering
	\small
	\caption{AUC(\%) of test-set performance on Avito, Avazu and Industrial datasets with various backbone and training strategy. * denotes one-pass learning. The results are averaged over 3 runs. Std $\le 0.1\%$. }
	\begin{tabular}{l|ccccc}
		\toprule
		Methods & Avazu & Avito & Avazu* & Avito* & Industrial*   \\
		\midrule
		DNN & 75.05 & 77.71 & 74.32 & 77.50 & 75.92 \\
		DCN & 74.99 & 77.66 & 74.30 & 77.58 & 75.99 \\
		PNN & 75.06 & 77.80 & 74.49 & 77.57 & n/a \\
		DeepFM & 75.24 & 77.73 & 74.69 & 77.50 & 76.02 \\
		\midrule
		DeepFM+KD & 75.41  & 78.01 & 74.83 & 77.54 & 76.18 \\
		DeepFM+RKD$_{l}$ & 75.34 & 77.83 & 74.90 & 77.58 & 76.10 \\
		DeepFM+SC & 75.36 & 77.78 & 74.85 & 77.53 & 76.14 \\
		\midrule
		DeepFM+CR$_{acc}$ & 75.63 & 78.33 & 74.98 & 77.70 & 76.25 \\
		DeepFM+RCR$_{auc}$ & 75.59 & 78.59 & 75.05 & \textbf{77.73} & 76.20   \\
		DeepFM+Both & \textbf{75.66} & \textbf{78.62} & \textbf{75.14} & 77.70 & \textbf{76.32} \\
		\bottomrule
	\end{tabular}
	
	\label{tab:ctr_prediction_results}
\end{table}

\begin{figure*}
	\centering
	
	\begin{subfigure}{0.27\textwidth}
		\includegraphics[width=\textwidth]{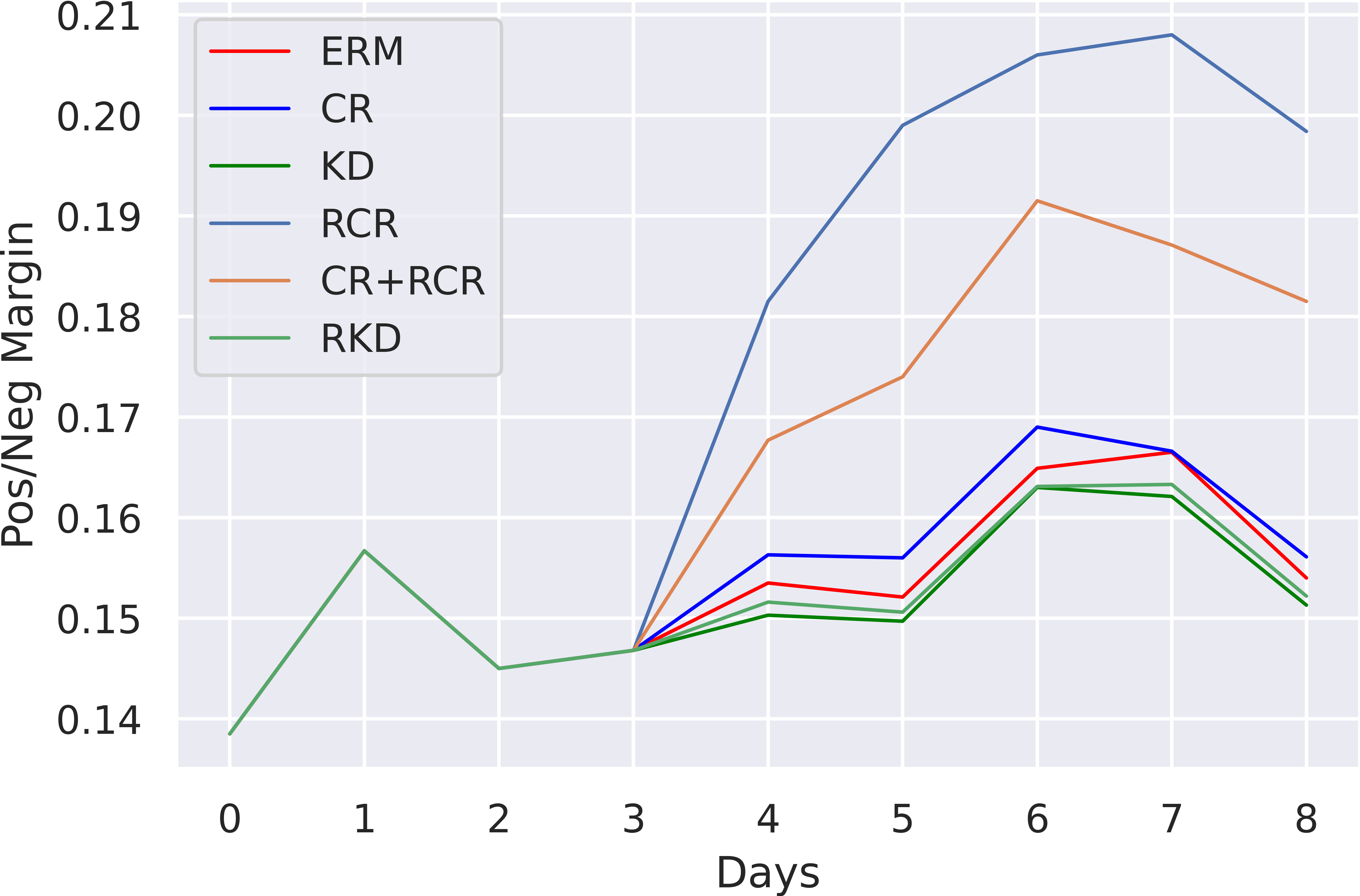}
		\caption{Sample Margin}
		\label{fig:sample_margin}	
	\end{subfigure}
	\quad
	\begin{subfigure}{0.27\textwidth}
		\includegraphics[width=\textwidth]{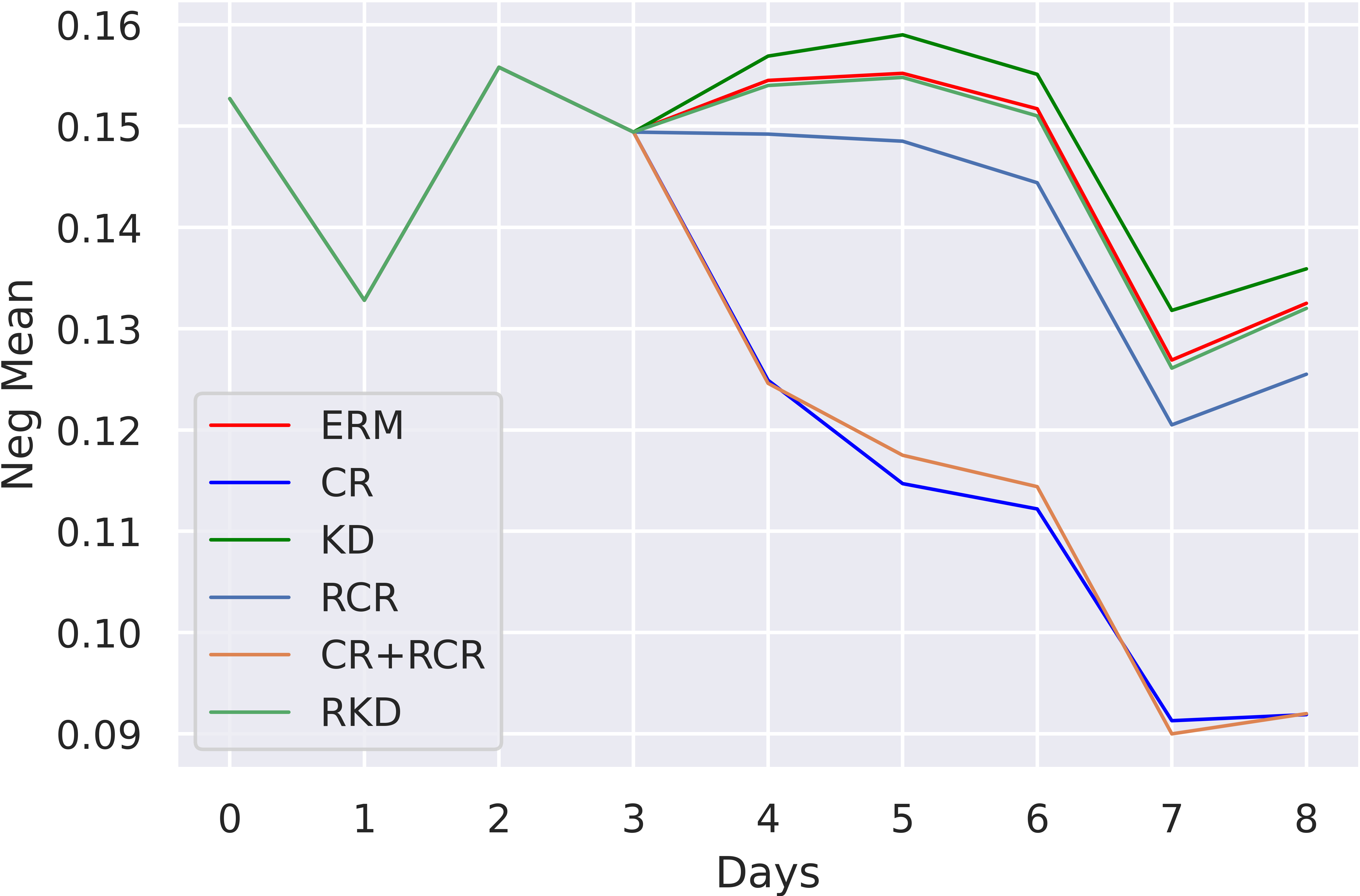}
		\caption{Negative Mean}
		\label{fig:neg_mean}
	\end{subfigure}
	\quad
	\begin{subfigure}{0.27\textwidth}
		\includegraphics[width=\textwidth]{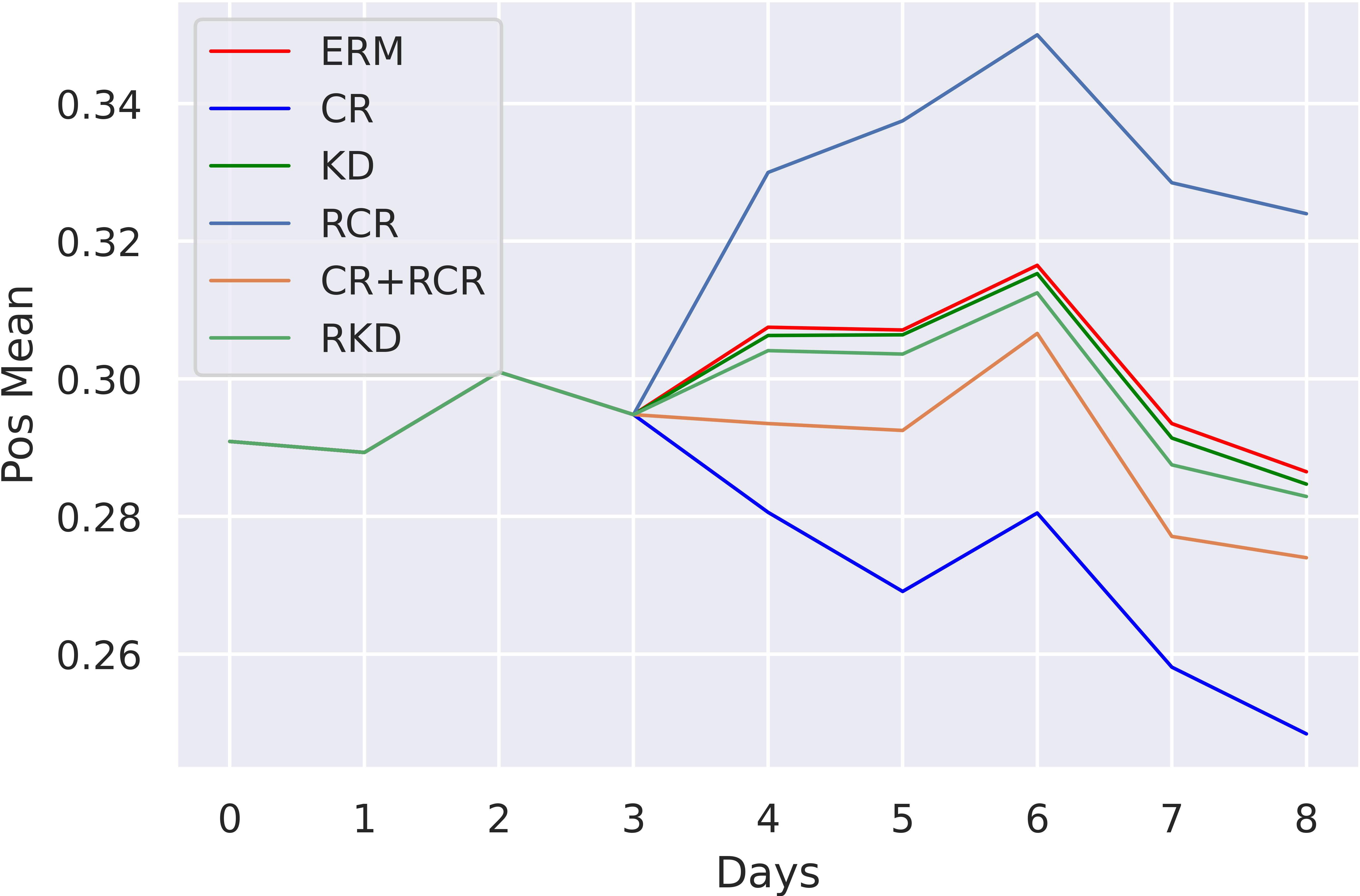}
		\caption{Positive Mean}
		\label{fig:pos_mean}
	\end{subfigure}
	
	\caption{Sample margin, prediction mean of negative  and positive samples on Avazu in one-pass setting.}
	\label{fig:margin}
\end{figure*}

\section{Experiments on CTR prediction}
We evaluate our methods on Industrial, Avazu and Avito datasets with various controllable setting on CTR prediction. Our core algorithm is easy to implement with various machine learning platform. For industrial datasets, we develop it with TensorFlow while we conduct experiments with PyTorch implementation for public datasets. All of our experiments are conducted on one P40 GPU for public datasets and 8 A100 GPUs for industrial dataset respectively.
	
\par \textbf{Datasets.} We perform our experiments on three datasets with two training setting.  Industrial search ads dataset contains 59 numerical and categorical feature fields. All of the fields data are discretized and transformed into sparse anonymous features. This dataset has more than ten billion instances range over one month with hundreds millions of active users and items. Avazu is display recommendation dataset released on Kaggle that contain 40428967 samples with 22 feature fields. Avito is also released as ads click datasets on Kaggle containing 190107687 samples but only 16 feature fields.  We construct public datasets by split into training/validation/test set by timestamp where the samples of last day is set for testing and penultimate day's data is set for validation and others are set for training. To split industrial dataset, we use traffic samples of previous 15 days  as training set and the last day as test set. We summarize statistic of datasets in Table \ref{tab:ctr_dataset} .

\par \textbf{Experiments setup}.  In real-world application, the prediction naturally influence the impression of items (i.e.  items that  have high confidence are more prone to be exposed to users.) and multiples times for training will cause severe \textit{over-fitting} issues and much more computation cost. Thus we adopt two various settings for evaluate our methods. The details of configuration are summarized as follows: (1) \textbf{One-Pass Setting:} we adopt one-pass training strategy to imitate \textit{online-learning} in a cycle  serving-and-training process with constant daily and minute-level data. For industrial dataset, it's common setting for evaluating performance of our methods. However, for public datasets, all of them only contain item and user features without any information of the deployed model. To overcome, we first train a one-pass model on $T$-days training set and then launch the cycling serving-and-training process with followed $\hat{T}$-days data (e.g. serve on $T+1$-th day data to get predictions as $f_{online}(x)$ then train with our method). We denote our experiments on industrial dataset only adopt one-pass training strategy. (2) \textbf{Standard Setting:} standard supervised learning. In this sense, all of our experiments train on training set with multiple epochs until the validation loss convergences. We use the outputs of previous epoch as the supervised signal for our proposed method.

\par \textbf{Baselines.}  Though our main motivation of our work is to utilize the confidence of online deployed model on target items, we still include several baselines under standard supervised learning and one-pass learning in order to benchmark state-of-the-art results. The simplest approach is (1) ERM: we train our networks with binary cross-entropy loss under two various settings. For majority of real-world recommendation and ads system, ctr prediction models are preferred to be trained with ERM; (2)  various commonly adopted CTR prediction network architectures designed for recommendation and ads system, DNN, PNN\cite{qu2016product}, DCN\cite{wang2017deep}, DeepFM\cite{guo2017deepfm}; (3) SC\cite{cai2022reloop} integrates self-correction module into CTR prediction networks. Together training with ERM, it achieves state-of-the-art results on multiple CTR prediction datasets with minimal computation cost.  (4) Knowledge distillation methods: since dark knowledge can induce useful gradients for model compression, we also adapt KD\cite{hinton2015distilling} and RKD\cite{park2019relational} to our experimental setting. In this paper, we modify the feature-based RKD to logit-based method for aligning inter-sample distance of the logits output of the base model and current model.
For all loss function we tune their loss balance term $\lambda$ ranging from 0.1 to 2.0. We select the best results in our experiments. The best $\lambda_{CR_{acc}}$ and $\lambda_{RCR_{auc}}$ is [0.4, 0.5] for public datasets and [0.5, 1.0] for industrial dataset.

\par \textbf{Main Results.} Table \ref{tab:ctr_prediction_results} compare the test-set AUC of our method on Click-Through-Rate prediction task. On Table \ref{tab:ctr_prediction_results}, we first investigate the improvement brought by different feature interaction methods. We observe that PNN achieve best performance with marginal improvement on standard supervised learning setting but fails compared to DeepFM and DCN on one-pass setting. For convenience, we adopt DeepFM as backbone for our experiments. We can observe that the propose method CR outperforms all baselines no matter which setting is adopted.  For standard supervised learning, it is also  striking to see that on Avazu and Avito, our proposed CR and RCR both can outperform baselines by a large margin after trained with multiple epochs. We denote 0.1\% improvement of AUC on Avazu and Avito is significant. For one-pass learning, we still observe that our proposed methods outperforms the backbone model but the margin is smaller than standard setting.  It's because one-pass learning may not completely fit on the two public datasets. For industrial dataset, we carefully tune our proposed method with DeepFM due to its succinct implementation. Not surprisingly, it works as well. Compared to vanilla distillation, our methods improve 0.25/0.61/0.31/0.16/0.14\% of the AUC respectively. 
\par \textbf{Inter-class margin Visualization.} In Figure \ref{fig:margin}, we show how sample margin, prediction mean value of positive and negative samples vary along time. The relational confidence ranking loss outperforms all the other method by a large margin. We can observe RCR both decrease the negative mean and increase the positive mean in Figure \ref{fig:neg_mean} and \ref{fig:pos_mean} leading to best bipartite ranking performance among all baseline methods in Table \ref{tab:ctr_prediction_results}. We find CR both decreases negative and positive mean resulting in marginal improvement on sample margin. We demonstrate it's because CTR prediction dataset is usually dominated by negative samples and our loss function tends to depress the negative prediction. In Figure \ref{fig:margin}, we find KD and RKD$_l$ give more smooth curve compared to our methods which may constrain the model's learning ability.

\begin{table}[t]
	\centering
	\caption{Results of click-through rate improvement on a 5-day online A/B experiment.}
	\begin{tabular}{ccccc|c}
		\toprule
		Day 1 & Day 2 & Day 3 & Day 4 & Day 5 & Average \\
		\midrule
		+1.62\% & +1.73\% & +1.90\% & +1.83\% & +1.67\% & 1.75\% \\ 
		\bottomrule
	\end{tabular}
	
	\label{tab:A/B_results}
\end{table}

\section{Online A/B Experiments}
Our architecture comprises of two parts: (1) In online ad serving platform, we additionally collect the outputs $y_{online}$ of $f(x; \theta_{online})$ as a online deployed prediction into our training data which directly decides which item will be exposed; (2) impose our proposed ranking-based loss to encourage the network to learn better than the online deployed one in both \textit{retraining} and \textit{online-learning} stage. 
\par \textbf{Online A/B results.} Additional to offline experiments, we conduct online experiments on A/B platform from 2022-8-15 to 2022-8-19. Our online A/B test experiments split active users into two groups. The first group is served by the recommendation results generated by current main model while the second is served by Confidence Ranking results. As shown in table \ref{tab:A/B_results}, We observe averaged 1.75\% improvement on CTR and apply it to serve main traffic in our system.

\section{Conclusion }
From the perspective of real-world application, we identify the problem of learning model for better generalization on \textit{retraining} and \textit{online-learning} stage compared to online deployed model. To address this problem, we propose a loss framework, named as Confidence Ranking, which compares various models' output predictions for maximizing the surrogate metric score. We extend this method to rank accuracy and Auc in CTR prediction. Our theoretical and experimental analysis shows that our method can effectively improve the results compared to cross-entropy optimization and distillation.

\newpage
\bibliographystyle{ACM-Reference-Format}
\bibliography{main.bib}


\begin{thebibliography}{13}


\ifx \showCODEN    \undefined \def \showCODEN     #1{\unskip}     \fi
\ifx \showDOI      \undefined \def \showDOI       #1{#1}\fi
\ifx \showISBNx    \undefined \def \showISBNx     #1{\unskip}     \fi
\ifx \showISBNxiii \undefined \def \showISBNxiii  #1{\unskip}     \fi
\ifx \showISSN     \undefined \def \showISSN      #1{\unskip}     \fi
\ifx \showLCCN     \undefined \def \showLCCN      #1{\unskip}     \fi
\ifx \shownote     \undefined \def \shownote      #1{#1}          \fi
\ifx \showarticletitle \undefined \def \showarticletitle #1{#1}   \fi
\ifx \showURL      \undefined \def \showURL       {\relax}        \fi
\providecommand\bibfield[2]{#2}
\providecommand\bibinfo[2]{#2}
\providecommand\natexlab[1]{#1}
\providecommand\showeprint[2][]{arXiv:#2}

\bibitem[Cai et~al\mbox{.}(2022)]%
        {cai2022reloop}
\bibfield{author}{\bibinfo{person}{Guohao Cai}, \bibinfo{person}{Jieming Zhu},
  \bibinfo{person}{Quanyu Dai}, \bibinfo{person}{Zhenhua Dong},
  \bibinfo{person}{Xiuqiang He}, \bibinfo{person}{Ruiming Tang}, {and}
  \bibinfo{person}{Rui Zhang}.} \bibinfo{year}{2022}\natexlab{}.
\newblock \showarticletitle{ReLoop: A Self-Correction Continual Learning Loop
  for Recommender Systems}.
\newblock \bibinfo{journal}{\emph{arXiv preprint arXiv:2204.11165}}
  (\bibinfo{year}{2022}).
\newblock


\bibitem[Dao et~al\mbox{.}(2021)]%
        {dao2021knowledge}
\bibfield{author}{\bibinfo{person}{Tri Dao}, \bibinfo{person}{Govinda~M
  Kamath}, \bibinfo{person}{Vasilis Syrgkanis}, {and} \bibinfo{person}{Lester
  Mackey}.} \bibinfo{year}{2021}\natexlab{}.
\newblock \showarticletitle{Knowledge Distillation As Semiparametric
  Inference}.
\newblock \bibinfo{journal}{\emph{arXiv preprint arXiv:2104.09732}}
  (\bibinfo{year}{2021}).
\newblock


\bibitem[Freund et~al\mbox{.}(2003)]%
        {freund2003efficient}
\bibfield{author}{\bibinfo{person}{Yoav Freund}, \bibinfo{person}{Raj Iyer},
  \bibinfo{person}{Robert~E Schapire}, {and} \bibinfo{person}{Yoram Singer}.}
  \bibinfo{year}{2003}\natexlab{}.
\newblock \showarticletitle{An efficient boosting algorithm for combining
  preferences}.
\newblock \bibinfo{journal}{\emph{Journal of machine learning research}}
  \bibinfo{volume}{4}, \bibinfo{number}{Nov} (\bibinfo{year}{2003}),
  \bibinfo{pages}{933--969}.
\newblock


\bibitem[Gneiting and Raftery(2007)]%
        {gneiting2007strictly}
\bibfield{author}{\bibinfo{person}{Tilmann Gneiting} {and}
  \bibinfo{person}{Adrian~E Raftery}.} \bibinfo{year}{2007}\natexlab{}.
\newblock \showarticletitle{Strictly proper scoring rules, prediction, and
  estimation}.
\newblock \bibinfo{journal}{\emph{Journal of the American statistical
  Association}} \bibinfo{volume}{102}, \bibinfo{number}{477}
  (\bibinfo{year}{2007}), \bibinfo{pages}{359--378}.
\newblock


\bibitem[Guo et~al\mbox{.}(2017)]%
        {guo2017deepfm}
\bibfield{author}{\bibinfo{person}{Huifeng Guo}, \bibinfo{person}{Ruiming
  Tang}, \bibinfo{person}{Yunming Ye}, \bibinfo{person}{Zhenguo Li}, {and}
  \bibinfo{person}{Xiuqiang He}.} \bibinfo{year}{2017}\natexlab{}.
\newblock \showarticletitle{DeepFM: a factorization-machine based neural
  network for CTR prediction}.
\newblock \bibinfo{journal}{\emph{arXiv preprint arXiv:1703.04247}}
  (\bibinfo{year}{2017}).
\newblock


\bibitem[Hinton et~al\mbox{.}(2015)]%
        {hinton2015distilling}
\bibfield{author}{\bibinfo{person}{Geoffrey Hinton}, \bibinfo{person}{Oriol
  Vinyals}, \bibinfo{person}{Jeff Dean}, {et~al\mbox{.}}}
  \bibinfo{year}{2015}\natexlab{}.
\newblock \showarticletitle{Distilling the knowledge in a neural network}.
\newblock \bibinfo{journal}{\emph{arXiv preprint arXiv:1503.02531}}
  \bibinfo{volume}{2}, \bibinfo{number}{7} (\bibinfo{year}{2015}).
\newblock


\bibitem[Hofst{\"a}tter et~al\mbox{.}(2020)]%
        {hofstatter2020improving}
\bibfield{author}{\bibinfo{person}{Sebastian Hofst{\"a}tter},
  \bibinfo{person}{Sophia Althammer}, \bibinfo{person}{Michael Schr{\"o}der},
  \bibinfo{person}{Mete Sertkan}, {and} \bibinfo{person}{Allan Hanbury}.}
  \bibinfo{year}{2020}\natexlab{}.
\newblock \showarticletitle{Improving efficient neural ranking models with
  cross-architecture knowledge distillation}.
\newblock \bibinfo{journal}{\emph{arXiv preprint arXiv:2010.02666}}
  (\bibinfo{year}{2020}).
\newblock


\bibitem[Kang et~al\mbox{.}(2021)]%
        {kang2021topology}
\bibfield{author}{\bibinfo{person}{SeongKu Kang}, \bibinfo{person}{Junyoung
  Hwang}, \bibinfo{person}{Wonbin Kweon}, {and} \bibinfo{person}{Hwanjo Yu}.}
  \bibinfo{year}{2021}\natexlab{}.
\newblock \showarticletitle{Topology distillation for recommender system}. In
  \bibinfo{booktitle}{\emph{Proceedings of the 27th ACM SIGKDD Conference on
  Knowledge Discovery \& Data Mining}}. \bibinfo{pages}{829--839}.
\newblock


\bibitem[Mohri et~al\mbox{.}(2018)]%
        {mohri2018foundations}
\bibfield{author}{\bibinfo{person}{Mehryar Mohri}, \bibinfo{person}{Afshin
  Rostamizadeh}, {and} \bibinfo{person}{Ameet Talwalkar}.}
  \bibinfo{year}{2018}\natexlab{}.
\newblock \bibinfo{booktitle}{\emph{Foundations of machine learning}}.
\newblock \bibinfo{publisher}{MIT press}.
\newblock


\bibitem[Park et~al\mbox{.}(2019)]%
        {park2019relational}
\bibfield{author}{\bibinfo{person}{Wonpyo Park}, \bibinfo{person}{Dongju Kim},
  \bibinfo{person}{Yan Lu}, {and} \bibinfo{person}{Minsu Cho}.}
  \bibinfo{year}{2019}\natexlab{}.
\newblock \showarticletitle{Relational knowledge distillation}. In
  \bibinfo{booktitle}{\emph{Proceedings of the IEEE/CVF Conference on Computer
  Vision and Pattern Recognition}}. \bibinfo{pages}{3967--3976}.
\newblock


\bibitem[{Qu} et~al\mbox{.}(2016)]%
        {qu2016product}
\bibfield{author}{\bibinfo{person}{Yanru {Qu}}, \bibinfo{person}{Han {Cai}},
  \bibinfo{person}{Kan {Ren}}, \bibinfo{person}{Weinan {Zhang}},
  \bibinfo{person}{Yong {Yu}}, \bibinfo{person}{Ying {Wen}}, {and}
  \bibinfo{person}{Jun {Wang}}.} \bibinfo{year}{2016}\natexlab{}.
\newblock \showarticletitle{Product-Based Neural Networks for User Response
  Prediction}. In \bibinfo{booktitle}{\emph{2016 IEEE 16th International
  Conference on Data Mining (ICDM)}}. \bibinfo{pages}{1149--1154}.
\newblock


\bibitem[Tang and Wang(2018)]%
        {tang2018ranking}
\bibfield{author}{\bibinfo{person}{Jiaxi Tang} {and} \bibinfo{person}{Ke
  Wang}.} \bibinfo{year}{2018}\natexlab{}.
\newblock \showarticletitle{Ranking distillation: Learning compact ranking
  models with high performance for recommender system}. In
  \bibinfo{booktitle}{\emph{Proceedings of the 24th ACM SIGKDD international
  conference on knowledge discovery \& data mining}}.
  \bibinfo{pages}{2289--2298}.
\newblock


\bibitem[{Wang} et~al\mbox{.}(2017)]%
        {wang2017deep}
\bibfield{author}{\bibinfo{person}{Ruoxi {Wang}}, \bibinfo{person}{Bin {Fu}},
  \bibinfo{person}{Gang {Fu}}, {and} \bibinfo{person}{Mingliang {Wang}}.}
  \bibinfo{year}{2017}\natexlab{}.
\newblock \showarticletitle{Deep \& Cross Network for Ad Click Predictions}. In
  \bibinfo{booktitle}{\emph{Proceedings of the ADKDD'17}}. \bibinfo{pages}{12}.
\newblock


\end{thebibliography}

\end{document}